\documentclass[letterpaper,conference]{IEEEtran}
\IEEEoverridecommandlockouts

\usepackage[utf8]{inputenc}
\usepackage{cite}
\usepackage{amsmath,amssymb,esint}
\usepackage{graphicx,xcolor}
\usepackage[super]{nth}
\usepackage{psfrag}
\usepackage{pbox,ragged2e}
\usepackage{balance}
\usepackage{placeins}
\usepackage[weather,misc,geometry,clock,electronic,alpine]{ifsym}

\usepackage[caption=false,font=footnotesize]{subfig}
\usepackage{textcomp}
\usepackage{gensymb}
\usepackage{dsfont}
\usepackage{bbm}
\usepackage{fontawesome} 

\newcommand{\f}[2]{\frac{#1}{#2}}

\newcommand{\Tr}{^\text{T}}

\DeclareMathOperator*{\argmax}{arg\,max}

  \newcommand{\p}{\mathbf{p}}
  \newcommand{\e}{\mathbf{e}}
\renewcommand{\d}{{\mathbf{d}}}

  \newcommand{\IndFunc}{\mathds{1}}

  \newcommand{\UB}{_\text{\,UMVUE}}
\renewcommand{\S}{\Delta}
  \newcommand{\SA}{\widetilde\S}
	\newcommand{\SN}{T}
	\newcommand{\SNA}{\tilde\SN}
	\newcommand{\err}{n}
\renewcommand{\dh}{\hat{d}}
  \newcommand{\eh}{\hat{\epsilon}}
  \newcommand{\synNL}{^{\,\text{(sync)}}}
  \newcommand{\asyNL}{^{\,\text{(asyn)}}}
  \newcommand{\syn}{^{\,\text{(sync,gen)}}}
  \newcommand{\asy}{^{\,\text{(asyn,gen)}}}

  \newcommand{\cd}{c\hspace{0.2mm}}
  
\newcommand{\Nk}[1]{_{\text{#1},k}}

\newcommand{\EVSymb}{\mathbb{E}}

\title{
Inter-Node Distance Estimation from Multipath Delay Differences of Channels to Observer Nodes
\thanks{This work was partially supported by the Commission for Technology and Innovation CTI, Switzerland and conducted in cooperation with Schindler Aufz\"uge AG.}}
\author{%
\IEEEauthorblockN{Gregor Dumphart, Marc Kuhn, and Armin Wittneben} 
\IEEEauthorblockA{Communication Technology Laboratory\\ ETH Zurich, Switzerland\\
Email: \{dumphart,  kuhn, wittneben\}@nari.ee.ethz.ch}
\and
\IEEEauthorblockN{Florian Tr\"osch} 
\IEEEauthorblockA{The PORT Technology\\ Schindler Aufz\"uge AG, Switzerland\\
Email: florian.troesch@schindler.com}}
\begin{document}

\maketitle

\begin{abstract}
We study the estimation of distance $d$ between two wireless nodes by means of their wideband channels to a \textit{third} node, called observer. 
The motivating principle is that the channel impulse responses are similar for small $d$ and drift apart when $d$ increases.
Following this idea we propose specific distance estimators based on the differences of path delays of the extractable multipath components. 
In particular, we derive such estimators for rich multipath environments and various important cases: with and without clock synchronization as well as errors on the extracted path delays (e.g. due to limited bandwidth). 
The estimators readily support (and benefit from) the presence of multiple observers.
We present an error analysis and, using ray tracing in an exemplary indoor environment, show that the estimators perform well in realistic conditions. We describe possible  localization applications of the proposed scheme and highlight its major advantages: it requires neither precise synchronization nor line-of-sight connection. This could make wireless user tracking feasible in dynamic indoor settings.
\end{abstract}
\section{Introduction}\label{sec:intro}
Most proposals for wireless localization systems rely on distance estimates to fixed infrastructure nodes (anchors) to determine the position of a mobile node \cite{BuehrerPIEEE2018}, e.g. via trilateration. Cooperative network localization furthermore employs the distances between different mobile nodes \cite{BuehrerPIEEE2018,LiJSAC2015,LiuTSP2018,MazuelasTSP2018}. 
A simple way to obtain such inter-node distance estimates is from the received signal strength (RSS) but the resultant accuracy is usually very poor due to shadowing, small-scale fading, and antenna patterns \cite{SchultenVTC2019}. A much more sophisticated method measures the time-of-arrival (TOA) with wideband signaling and a round-trip protocol for synchronization \cite{BuehrerPIEEE2018,DardariPIEEE2009}.

TOA-based localization schemes require involved hardware at both ends and suffer from synchronization errors and processing delays \cite{WymeerschPIEEE2009,DardariPIEEE2009,AlaviCL2006,JourdanTAES2008}.
Yet the main problem is ensuring a sufficient number of anchors in line of sight (LOS) to all relevant mobile positions \cite{WitrisalSPM2016}. TOA thus exhibits a large relative error at short distances and is not well-suited for dense and crowded settings such as lobbies, metro stations, access gates, and large events. These however entail important use cases (e.g., see \cite{Gani2016}).
The related time difference of arrival (TDOA) scheme 
does not offer a solution because it suffers the same non-LOS problem as TOA, requires precise synchronization between the anchors (which hinders their distribution and coverage), and cannot be used for inter-mobile distance estimation.

In this paper we propose and study an alternative paradigm for inter-node distance estimation (which, to the best of our knowledge, has not received attention so far) with the aim of alleviating the outlined problems of wireless localization systems. To begin with, we abandon the notion that an estimate of the distance $d$ between two nodes A and B should be based on a direct measurement such as the TOA or RSS between them. 
Instead, we consider the presence of \textit{another} node, 
henceforth called \textit{observer} node. We furthermore assume the availability of the channel impulse response (CIR) $h_\text{A}(\tau)$ of the channel between node A and the observer as well as CIR $h_\text{B}(\tau)$ between node B and the observer. The CIRs can be obtained via channel estimation at the observer after transmitting wideband training sequences at A and B \cite{MolischPIEEE2009}. The basic setup is shown in Fig.~\ref{fig:AppBeacon}.
The starting point of this paper is the observation that the CIRs $h_\text{A}(\tau)$ and $h_\text{B}(\tau)$ are similar for small $d$ and that this \textit{similarity vanishes} steadily with increasing $d$.
\newcommand{\appHSpace}{\hspace{3mm}}%
\newcommand{\appFigWidth}{.75\columnwidth}%
\begin{figure}[!b]
    \vspace{-.8mm}
    \subfloat[Single observer]{
        \appHSpace
        \centering 
        \includegraphics[width=\appFigWidth,trim=0 86 0 28mm,clip=true]{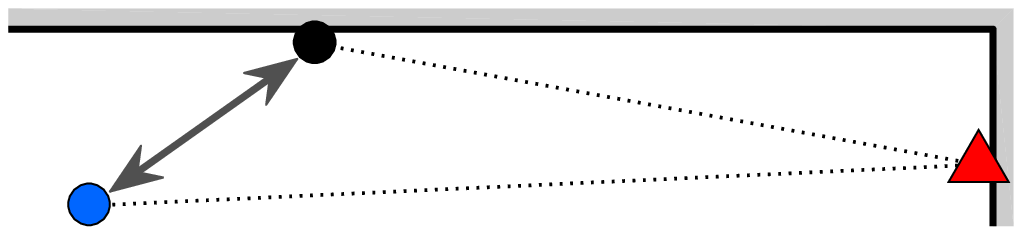}
        \label{fig:AppBeacon}}
        \put(-147,16){$d$}
        \put(-21,9){\footnotesize{\textcolor[rgb]{.8,0,0}{\begin{tabular}{c}observer\\node\end{tabular}}}}
        \put(-142,35.0){\footnotesize{node A}}
        \put(-90,35.0){\footnotesize{\textcolor[rgb]{0.3,0.3,0.3}{(walls of indoor environment)}}}
        \put(-188.0,3.5){\footnotesize\textcolor[rgb]{0,.2,.8}{node B}}
        \put(-72,19.0){\footnotesize{$h_\text{A}(\tau)$}}
        \put(-113,9){\textcolor[rgb]{0,.2,.8}{\footnotesize{$h_\text{B}(\tau)$}}}
    \ \\[2.4mm]
		\subfloat[Improving the scheme with additional observer nodes]{
        \centering
        \appHSpace
        \includegraphics[width=\appFigWidth,trim=0 28mm 0 19mm,clip=true]{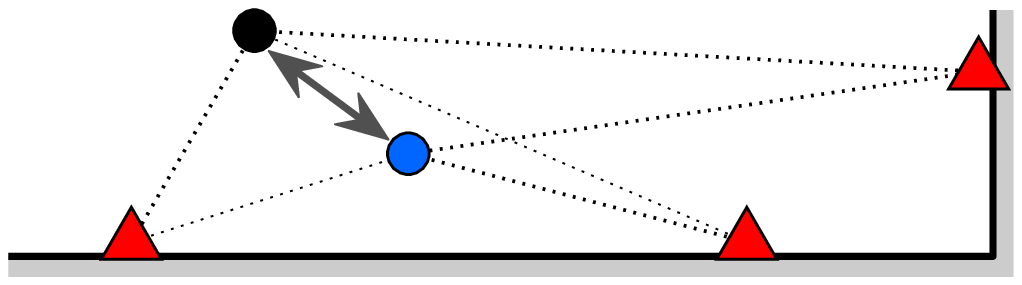}
        \put(-126,16.0){$d$}
        \put(-160,33){\footnotesize{node A}}
        \put(-114.2,6.2){\footnotesize\textcolor[rgb]{0,.2,.8}{node B}}
        \put(-18.5,18){\footnotesize{\textcolor[rgb]{.8,0,0}{\begin{tabular}{c}multiple\\observer\\nodes\end{tabular}}}}
        \label{fig:AppCooperative}}
    \ \\[2.4mm]
		\subfloat[Mobile observers, e.g. for network localization]{
        \centering
        \appHSpace
        \includegraphics[width=\appFigWidth,trim=0 28mm 0 17mm,clip=true]{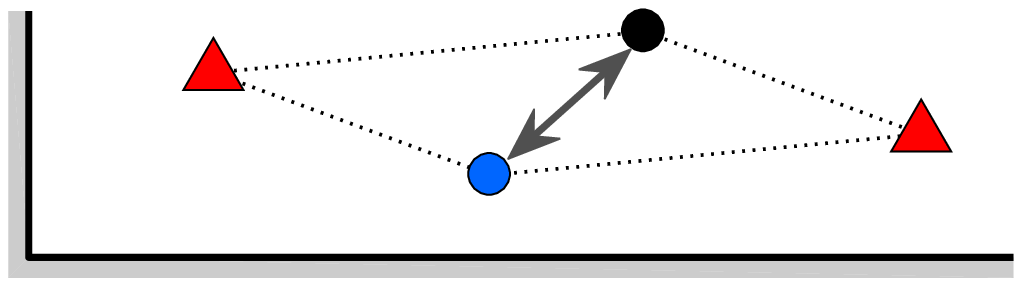}
        \put(-74.2,20.7){$d$}
        \put(-62.8,36.5){\footnotesize{node A}}
        \put(-100.0,5.2){\footnotesize\textcolor[rgb]{0,.2,.8}{node B}}
        \put(-23.5,18){\footnotesize{\textcolor[rgb]{.8,0,0}{\begin{tabular}{c}also mobiles\\can serve as\\observers\end{tabular}}}}
        \label{fig:AppNetwork}}
    \caption{Proposed scheme for estimation of the distance $d$ between two wireless nodes A and B in different possible setups. The estimation shall be based on the similarity of the  CIRs $h_\text{A}(\tau)$ and $h_\text{B}(\tau)$ to an observer node (or the similarity of all their respective CIRs to multiple observer nodes). The gray walls indicate indoor environments with rich multipath propagation.} 
    \label{fig:Apps}
\end{figure}
A good metric for the similarity between the CIRs could give rise to an accurate estimate $\hat{d}$ as a function of this metric,
with the prospect of particularly good performance at short distances (due to the focus on local channel variations) and no requirements for LOS connections.

From an application perspective (details follow in Sec.~\ref{sec:sota}), the setup in Fig.~\ref{fig:AppBeacon} evaluates proximity to a stationary node, e.g. some point of interest. If distance estimates to multiple stationary nodes at known positions are obtained this way, trilateration of the mobile position can be performed.
Fig.~\ref{fig:AppCooperative} and \ref{fig:AppNetwork} are concerned with inter-mobile distances, e.g. for network localization.
They also show the possibility of using multiple observers, which can be fixed infrastructure (\ref{fig:AppCooperative}) or other mobiles (\ref{fig:AppNetwork}).

To tap the great potential of the proposed paradigm, the remainder of the paper focuses on a specific realization that is based on the multipath delay structure of the CIRs. In Sec.~\ref{sec:estimators} we derive corresponding distance estimates, in particular for the practically important case of unsynchronized nodes and using extracted delays that are subject to errors. Thereby we employ estimation theory and assumptions regarding rich multipath propagation.
In Sec.~\ref{sec:eval} we evaluate performance and limitations under ideal conditions and in a more realistic indoor setting using ray tracing. Sec.~\ref{sec:sota} discusses the accuracy and technological opportunities of the scheme in the context of indoor localization. Sec.~\ref{sec:summary} then concludes the paper.


%

%
%
\section{Distance Estimates from Delay Differences}\label{sec:estimators}
\begin{figure}[!ht]
	\subfloat[Multipath propagation from nodes A and B to the observer]{
    \centering
    \includegraphics[height=4cm,trim=0 48 0 33,clip=true]{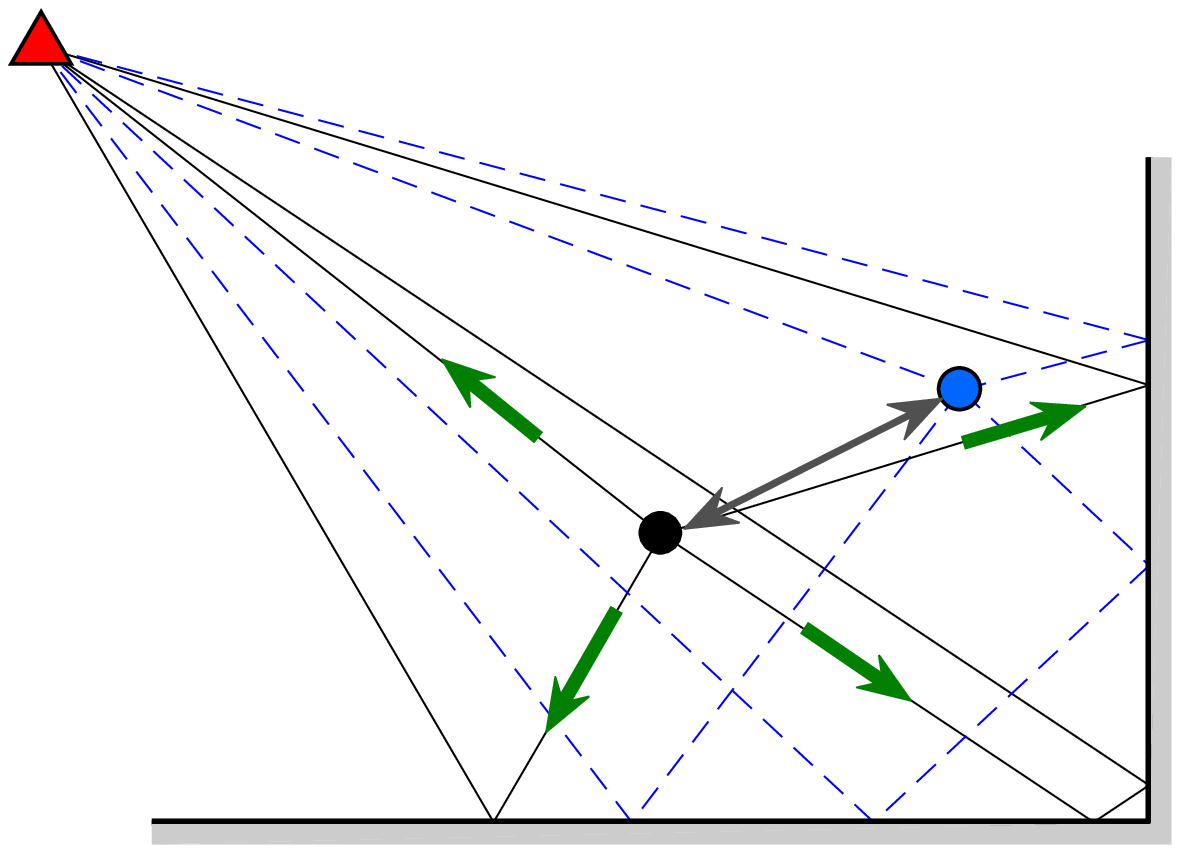}
		\put(-115.2,69.3){\textcolor[rgb]{1,1,1}{$\bullet$}}
		\put(-112.2,67.6){\textcolor[rgb]{1,1,1}{$\bullet$}}
		\put(-109.8,65.8){\textcolor[rgb]{1,1,1}{$\bullet$}}
    \put(-113.2,69.8){\textcolor[rgb]{0,.5,0}{\footnotesize$\e_1$}}
		\put(-93,22){\footnotesize\textcolor[rgb]{0,.5,0}{$\e_2$}}
		\put(-33.0,52.5){\footnotesize\textcolor[rgb]{0,.5,0}{$\e_3$}}
		\put(-65,17){\footnotesize\textcolor[rgb]{0,.5,0}{$\e_4$}}
		\put(-101.9,42.5){\textcolor[rgb]{1,1,1}{$\bullet$}}
		\put(-100.8,44.2){\footnotesize{$\p_\text{A}$}}
    \put(-60.7,65.1){\textcolor[rgb]{1,1,1}{$\bullet$}}
		\put(-59.0,68.5){\textcolor[rgb]{0,.2,.8}{\footnotesize{$\p_\text{B}$}}}
		\put(-72,57.0){\footnotesize{$d$}}
		\put(-145,108){\footnotesize{\textcolor[rgb]{.8,0,0}{observer node}}}
		\put(-5,30){\line(0,1){6}}
		\put(34.6,30){\line(0,1){6}}
		\put(-5,33){\line(1,0){39.6}}
		\put(8,36){$3\,\mathrm{m}$}
		\vspace{-3mm}
    \label{fig:InterUserPaths}}
  \ \\
  \subfloat[Seas][CIR $h_\text{A}(\tau)$ from node A to observer node;\\\hphantom{(b) }CIR $h_\text{B}(\tau)$ from node B to observer node]{
    \centering
    \psfrag{ht}{\raisebox{.5mm}{\hspace{-3mm}\footnotesize{$|h_*(\tau)|$}}}
    \includegraphics[width=\columnwidth,trim=19mm -12 16mm 5mm,clip=true]{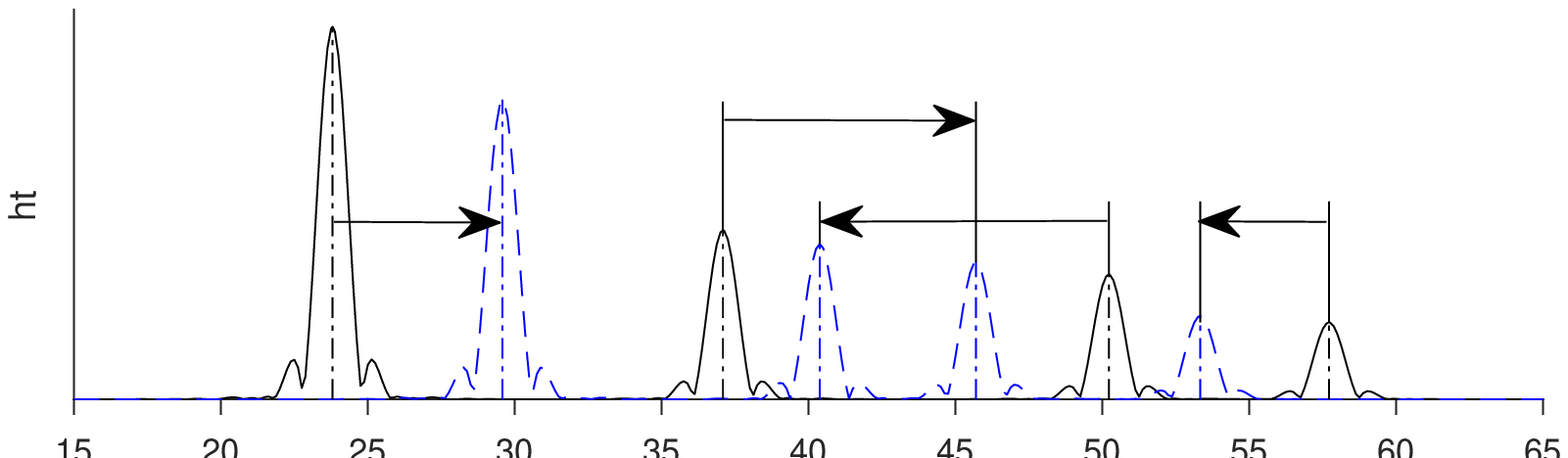}
    \put(-134.5,-2){\footnotesize{$\tau\,\mathrm{[ns]}$}}
    \put(-225.0,62){\footnotesize{$h_\text{A}(\tau)$}}
		\put(-175,68){\textcolor[rgb]{0,.2,.8}{\footnotesize{$h_\text{B}(\tau)$}}}
		\put(-192,48.5){\footnotesize{$\S_1$}}
		\put(-123,64.5){\footnotesize{$\S_2$}}
		\put(-112,48.5){\footnotesize{$\S_3$}}
		\put( -53,48.5){\footnotesize{$\S_4$}}
    \label{fig:InterUserSignals}}
	\caption{Concept of distance estimation between two nodes A and B by comparing (the path delays of) their wideband CIRs to an observer node. The upper plot depicts this approach in an indoor environment with two walls, also showing the significant propagation paths. The plot below shows the two corresponding CIRs (responses to raised-cosine pulses of $1\,\mathrm{GHz}$ bandwidth
) with $K = 4$ MPCs and illustrates the delay differences concept.}
	\label{fig:InterUserIntro}
  \vspace{-2.85mm}
\end{figure}

We consider the setup in Fig.~\ref{fig:InterUserIntro} with the nodes A and B with distance $d$ and an observer node located in a multipath propagation environment. We express $d = \|\d\|$ in terms of the displacement vector $\d = \p_\text{B} - \p_\text{A} \in \mathbb{R}^3$ from node A at position $\p_\text{A} \in \mathbb{R}^3$ to node B at $\p_\text{B} \in \mathbb{R}^3$. The unit vectors $\e_k \in \mathbb{R}^3$ denote the multipath directions of departure at $\p_\text{A}$.

Given $h_\text{A}(\tau)$ and $h_\text{B}(\tau)$ from nodes A and B to the observer, we want to determine $d$ by a comparison of the CIRs.
If those CIRs are estimated with large bandwidth, several multipath components (MPCs) are usually resolvable and can be extracted \cite{MolischPIEEE2009}. We consider only the subset of MPCs that occur in both CIRs (propagation paths that emerge from both $\p_\text{A}$ and $\p_\text{B}$ to the observer, cf. \cite{WitrisalSPM2016}) and that were successfully extracted from both. We denote $\tau\Nk{A}$ and $\tau\Nk{B}$ for the MPC path delays, whereby indexation $k = 1 \ldots K$ is such that delays of equal $k$ arise from the same propagation path\footnotemark{} (e.g. via the same reflector or scatterer).
\mbox{$K$ is the} number of MPCs that were extracted from both CIRs.

The node displacement causes delay differences
\begin{align}
\S_k &= \tau\Nk{B} - \tau\Nk{A} \ , &
k &= 1, \ldots, K
\label{eq:ShiftDef}
\end{align}
over equal propagation paths, as illustrated in Fig.~\ref{fig:InterUserSignals}.
The enabling fact for our approach is that all $\S_k$ are subject to the bounds\footnotemark{}
$-d \leq \cd\S_k \leq d$
due to propagation at the speed of light $c$.
Because of this geometric significance we consider $\S_k$ as key observable quantity for distance estimation: each value yields a lower bound $d \geq \cd|\S_k|$ on the distance.  With all observations considered we get 
$d \geq c \cdot \max_k |\S_k|$,
a tight bound whenever the direction of $\d$ is similar to $\e_k$ or $-\e_k$ for any $k$. This is highly probable when $K$ is large and the MPCs have diverse directions, which is characteristic for dense indoor or urban environments. In this case, we can compute an accurate distance estimate
\begin{align}
\dh\synNL = c \cdot \max_k |\S_k| .
\label{eq:dMLSync}
\end{align}

\footnotetext[1]{We note from Fig.~\ref{fig:InterUserSignals} that the association between the MPCs across the two CIRs (comprising the problem of finding the subset of common MPCs) is a non-trivial task; a nearest neighbor scheme will usually fail unless $d$ is very small. Such association problems however have been studied thoroughly, e.g., for a single temporal snapshot in \cite{DokmanicPNAS2013} and for temporal tracking in \cite[and references therein]{WitrisalSPM2016}. In this paper we assume perfect association and leave an evaluation of the cited methods in this context to future work.}%

\footnotetext[2]{To obtain these bounds formally, denote $\p_k \in \mathbb{R}^3$ for the $k$-th MPC virtual sink position, e.g., the observer position mirrored at the wall(s) of a reflection \cite{WitrisalSPM2016}. Write $\cd\S_k = \|\p_\text{B} - \p_k\| - \|\p_\text{A} - \p_k\|$ and, using $d = \|\p_\text{B} - \p_\text{A}\|$, obtain $\cd\S_k \geq -d$ and $\cd\S_k \leq d$ from the triangle inequality.}%

Measuring the values $\S_k$ however requires precise time synchronization between the two nodes (sub-$\mathrm{ns}$ precision) which can hardly be achieved with mobile consumer electronics.
An alternative is to consider asynchronous delay differences
\begin{align}
\SA_k = \S_k + \epsilon
\label{eq:ShiftDefAsyn}
\end{align}
as observations, subject to an \textit{unknown clock offset} $\epsilon$ (the same for all $k$). In this case estimation rule \eqref{eq:dMLSync} cannot be applied. Yet we can find a meaningful distance estimate 
by looking at the value range $\cd\epsilon - d \leq c\SA_k \leq \cd\epsilon + d$, again an interval of width $2d$. For large $K$ and diverse MPC directions
we expect $c\cdot \min_k \SA_k$ and $c\cdot \max_k \SA_k$ close to the lower and upper bounds, respectively. We are hence able to compute
a distance estimate from asynchronous observations
\begin{align}
\dh\asyNL &= \f{c}{2} \Big( \max_k \SA_k - \min_k \SA_k \Big) . 
\label{eq:dMLAsyn}
\end{align}

So far our approach has been heuristical and we like to formalize it by means of estimation theory. For this we need to establish statistics for the observations $\S_k$. We do so with the following \textit{assumptions} on the MPC directions $\e_k$.
\newcommand{\Ass}[1]{\uppercase\expandafter{\romannumeral#1}}
\begin{enumerate}
\item[\Ass{1}:] The MPC directions $\e_k$ are the same at $\p_\text{A}$ and $\p_\text{B}$.
\item[\Ass{2}:] $\e_k$ is random and all directions are equiprobable, i.e. $\e_k$ has uniform distribution 
on the 3D unit sphere. 
\item[\Ass{3}:] The directions $\e_k$ and $\e_l$ of different paths $k \neq l$ are statistically independent.
\end{enumerate}
By \Ass{1} we assume a locally constant MPC geometry.%
\footnote{In detail, assumption \Ass{1} is valid when $d$ is much smaller than the distances from $\p_\text{A}$ and $\p_\text{B}$ to the virtual sink of the MPC in question (cf. \cite{WitrisalSPM2016}).}
This is equivalent to a plane-wave approximation if the observer was transmitting and is supported by the example in Fig.~\ref{fig:InterUserPaths} to a large extend.
Therewith we can relate the delay differences to projections
$\cd\S_k = -\e_k\Tr {\bf d}$
of the displacement vector.%
\footnote{If the directions $\e_k$ were known, $\d$ could be determined from the linear system of $K$ equations $\cd\S_k = -\e_k\Tr {\bf d}$, but this would require specific knowledge about the environment as in multipath-assisted localization \cite{WitrisalSPM2016,LeitingerJSAC2015} and is not possible with our statistical description of the $\e_k$.} 
By a convenient property\footnotemark{} of uniformly distributed 3D unit vectors, the resultant observation statistic is the uniform distribution
\begin{align}
\cd\S_k \overset{\text{i.i.d.}}{\sim} \ \mathcal{U}(-d,+d) .
\label{eq:UniformShift}
\end{align}
\footnotetext{In detail, \eqref{eq:UniformShift} follows from projection $\cd\S_k = -\e_k\Tr {\bf d}$ and assumptions \Ass{2} and \Ass{3}. This holds because any projection of a uniformly distributed 3D unit vector has uniform distribution. This little known property of 3D space is a consequence of Archimedes' hat box theorem or, equivalently, the fact that the surface area of a spherical cap is proportional to its height. It has been employed in the wireless context in \cite[Eq.(12)]{DumphartPIMRC2016} where a short proof is given.}
We are now ready for an estimation-theoretic study of the proposed distance estimation scheme. In the following we summarize our key findings for four relevant cases. The derivations can be found in the appendix.

\subsection{Delays extracted without error; synchronous clocks}
We assume the delay differences $\S_k$ are available exactly as defined in \eqref{eq:ShiftDef}. This requires that (i) the delays $\tau\Nk{A}$ and $\tau\Nk{B}$ were extracted from the respective CIRs without error, e.g. by using a very large bandwidth, and (ii) the clocks of node A and B are perfectly synchronous. For this case and the assumed MPC statistics, we find that $\dh\synNL$ in \eqref{eq:dMLSync} is the maximum likelihood estimate (MLE) of $d$. It is an underestimate with probability $1$ (because any $\e_k$ hits the exact direction of ${\bf d}$ or $-{\bf d}$ with probability $0$) and the bias is $\EVSymb[\dh\synNL] - d = -\f{d}{K+1}$. A simple bias correction of \eqref{eq:dMLSync} leads to an unbiased estimate
\begin{align}
\dh\UB\synNL = \f{K+1}{K}\ c \cdot \max_k |\S_k| 
\label{eq:dUBSync}
\end{align}
which is in fact the uniform minimum-variance unbiased estimate (UMVUE) for this problem.

\subsection{Delays extracted without error; asynchronous clocks}
We consider the case where time synchronization is not established or required but asynchronous delay differences $\SA_k$ are available as defined in \eqref{eq:ShiftDefAsyn}. 
The estimate $\dh\asyNL$ in \eqref{eq:dMLAsyn} is the MLE for the assumed MPC statistics. The bias is $\EVSymb[\dh\asyNL] - d = -\f{2d}{K+1}$ and, therefrom, an unbiased estimate
\begin{align}
\dh\UB\asyNL = \f{K+1}{K-1}\cdot \f{c}{2} \Big( \max_k \SA_k - \min_k \SA_k \Big)
\label{eq:dUBAsyn}
\end{align}
is obtained, which is the UMVUE for this problem.

It is worth noting the associated clock offset estimate
\begin{align}
\eh\UB\asyNL &= \f{1}{2} \Big( \max_k \SA_k + \min_k \SA_k \Big)
\label{eq:eMLAsyn}
\end{align}
which could be useful by itself for distributed synchronization in dense multipath. It is both the MLE and the UMVUE.%
\footnote{It can be shown that \eqref{eq:dMLSync}, \eqref{eq:dMLAsyn}, \eqref{eq:eMLAsyn} are the MLE also for the respective 2D cases with analogous assumptions on $\e_k$. Instead of \eqref{eq:UniformShift}, this case features $f(\S_k|d) = \frac{c}{\pi} (d^2 - c^2\S_k^2)^{-1/2}$ as observation PDF. The details are omitted.}

\subsection{General case with synchronous clocks}
\label{sec:subErrSync}

When the path delays are measured with error (e.g., due to limited bandwidth), the distance estimates introduced so far might get distorted heavily: they are very susceptible to outliers since they regard only the maximum and minimum delay difference. It is thus sensible to include such errors in the statistical model and derive according distance estimates. 

We first consider the case of perfectly synchronous nodes but with observed delay differences
$\SN_k = \S_k + \err_k$
subject to random errors $\err_k$ (as a result of delay extraction errors). 
We assume that the distribution of $\err_k$ is known and furthermore that $\err_k$ and $\err_l$ are statistically independent for $k \neq l$. The resulting distance MLE is given by the optimization problem
\begin{align}
& \dh\syn
\in \argmax_d \,\f{1}{d^K} \prod_{k=1}^K I_k(\SN_k,d),
\label{eq:dMLSyncNoisy} \\
& I_k(\SN_k,d) = F_{\err_k}(\SN_k + d/c) - F_{\err_k}(\SN_k - d/c)
\label{eq:softInd}
\end{align}
where $F_{\err_k}$ is the cumulative distribution function (CDF) of $\err_k$.
Hence $I_k$ can be regarded as a soft indicator function of $\cd \SN_k \in [-d,d\,]$.
%
If the errors have Gaussian distribution\footnotemark{} $\err_k \sim \mathcal{N}(0,\sigma_k^2)$ we can use the $Q$-function to write
\begin{align}
I_k(\SN_k,d) &= Q\left(\f{\SN_k - d/c}{\sigma_k}\right) - Q\left(\f{\SN_k + d/c}{\sigma_k}\right).
\label{eq:softIndGaussian}
\end{align}

Estimate \eqref{eq:dMLSyncNoisy} is biased in general. This is seen by the example of errorless extraction: $\err_k \equiv 0$ results in $I_k(\SN_k,d) = \IndFunc_{[-d,d\,]}(\cd \SN_k)$ (the actual indicator function) and consequently \eqref{eq:dMLSyncNoisy} yields \eqref{eq:dMLSync} as a special case (the proof is straightforward) which we know is a biased estimate.

\footnotetext{A Gaussian error on the delay differences could be the result of the delays $\tau\Nk{A}$ and $\tau\Nk{B}$ being extracted subject to uncorrelated Gaussian errors. This model is suggested by \cite{Qi2003} for high SNR.}

\subsection{General case with asynchronous clocks}

Finally, we consider the case where erroneous asynchronous delay differences
$\SNA_k = \SN_k + \epsilon = \S_k + \err_k + \epsilon$
are observed, i.e. subject to a clock offset $\epsilon$ and an extraction error $\err_k$. The joint MLE of distance and clock offset is given by
\begin{align}
\left( \dh\asy , \eh\asy \right)
\in \argmax_{d,\epsilon} \,\f{1}{d^K} \prod_{k=1}^K I_k(\SNA_k - \epsilon,d) .
\label{eq:dMLAsynNoisy}
\end{align}

This distance estimate is biased in general (an unbiased estimate remains as an open problem). This is seen at the special case of errorless extraction, analogous to Sec.~\ref{sec:subErrSync}: for $\err_k \equiv 0$ it can be shown that \eqref{eq:dMLAsynNoisy} yields the biased \eqref{eq:dMLAsyn}.

\subsection{Technical aspects and comments}

A subtle but important aspect is that the presented estimators can \textit{incorporate multiple observer nodes} without further ado by simply considering the MPCs from nodes A and B to all observer nodes (use index $k$ on this set). The increased number of observations can improve performance considerably.

We never assumed or required knowledge of the observer positions, synchronization between observer(s) and the other two nodes, or synchronization among multiple observers as such circumstances would not even improve the scheme. These are the \textit{key complexity advantages} of our proposal.

Another fortunate aspect is that assumption \Ass{1} is almost superfluous because any MPC relevant to estimation (one with large $|\S_k|$) fulfills it quite naturally: large $|\S_k|$ corresponds to $\e_k$ being similar to the direction of $\d$ or $-\d$ and, thus, $\e_k$ hardly changes when moving by $\d$.

The properties of optimization problems \eqref{eq:dMLSyncNoisy} and \eqref{eq:dMLAsynNoisy} depend on the error CDFs $F_{\err_k}$. With Gaussian error statistics \eqref{eq:softIndGaussian} the problems are non-convex (because the $Q$-function is non-convex) yet very amenable: in all conducted experiments, the likelihood function was unimodal and the problems could be solved with very few iterations of a gradient-based solver.

\section{Performance Evaluation}\label{sec:eval}
This section discusses various sources of error and their effect on the accuracy of the proposed distance estimates.

\subsection{Impact of unknown MPC directions}
\label{sec:EvalUnknown}
An important source of error are the unknown $\e_k$ which determine the observed delay differences. Because of the mathematical simplicity of the observation statistics \eqref{eq:UniformShift}, the resulting estimation error statistics can be described in closed form (for derivations see the appendix). 

The root-mean-squared error (RMSE) of the synchronous-case UMVUE \eqref{eq:dUBSync} is given by its standard deviation
\begin{align}
\mathrm{std}\Big[ \dh\UB\synNL \Big] = \f{d}{\sqrt{K(K+2)}}
\label{eq:stdUBSync}
\end{align}
while the RMSE of asynchronous estimates \eqref{eq:dUBAsyn} and \eqref{eq:eMLAsyn} is characterized by the large-$K$ approximations
\begin{align}
\mathrm{std}\Big[ \dh\UB\asyNL \Big] 
&\approx \f{d}{K-1}\ \sqrt{\f{2K}{K+2}}
\label{eq:stdUBAsyn} \ , \\
\mathrm{std}\Big[ \eh\UB\asyNL \Big]
&\approx \f{d/c}{K+1}\ \sqrt{\f{2K}{K+2}}
\label{eq:stdeUBAsyn}
\end{align}
which are accurate for about $K \geq 5$. All errors are proportional to $d/K$ asymptotically. In other words, the error of distance estimation based on unknown MPC directions increases linearly with distance.

For reliable estimation of the distance between synchronous nodes, a single $\e_k$ similar to the direction of \textit{either} $\d$ or $-\d$ suffices. The asynchronous case, in contrary, requires $\e_k$ similar to the directions of \textit{both} $\d$ and $-\d$ to occur and is thus more reliant on diverse $\e_k$. This fundamental difference is due to the unknown clock offset $\epsilon$ and is apparent when comparing \eqref{eq:dMLSync} to \eqref{eq:dMLAsyn}. The performance difference can be quantified as $\mathrm{std}[ \dh\UB\asyNL ] \approx \sqrt{2}\ \mathrm{std}[ \dh\UB\synNL ]$ under our assumptions.

\subsection{Impact of delay extraction errors}
\label{eq:EvalExtract}
\begin{figure}[!ht]
	\centering
	\subfloat[Impact of $K$; error standard deviation fixed to $\cd\sigma/d = 0.5$]{\shortstack[l]{
    \psfrag{K}{\hspace{-19mm}\raisebox{-.2mm}{\scriptsize number of common detected MPCs $K$}}
		\psfrag{rb}{\hspace{-5.1mm}\raisebox{0mm}{\scriptsize relative bias}}
		\includegraphics[width=\columnwidth,trim=4mm  0mm 10mm 2mm,clip=true]{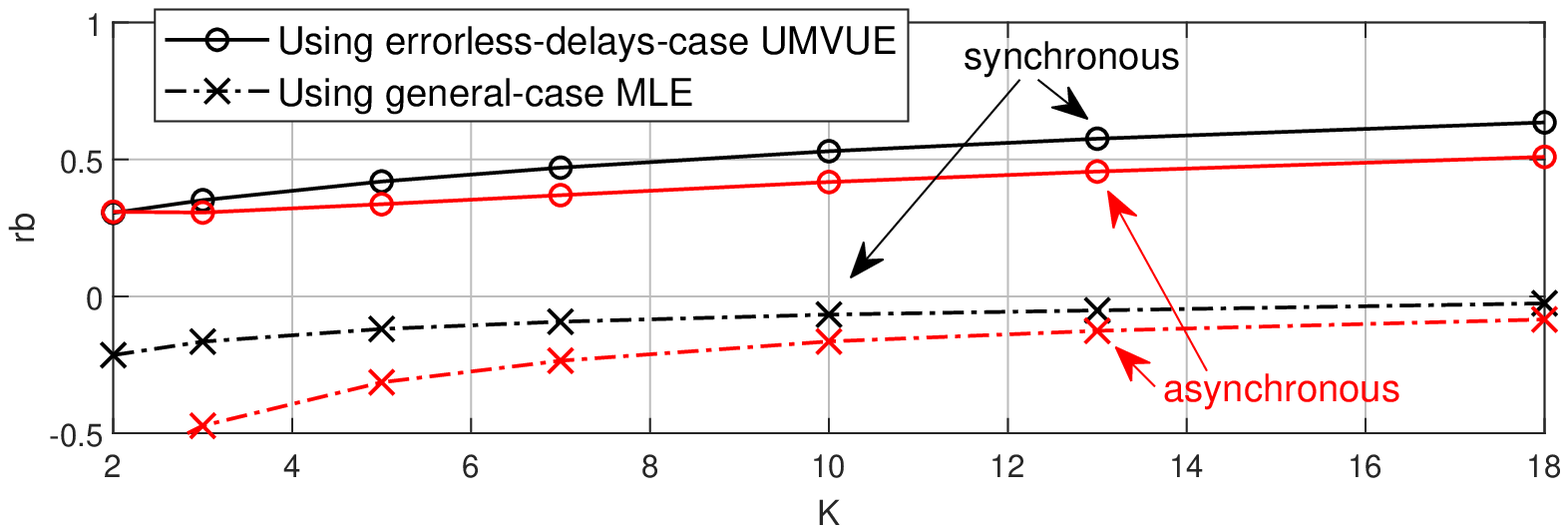} \\
    \psfrag{K}{\hspace{-19mm}\raisebox{-.2mm}{\scriptsize number of common detected MPCs $K$}}
		\psfrag{rr}{\hspace{-7.5mm}\raisebox{0mm}{\scriptsize relative RMSE}}
		\includegraphics[width=\columnwidth,trim=4mm -3mm 10mm 2mm,clip=true]{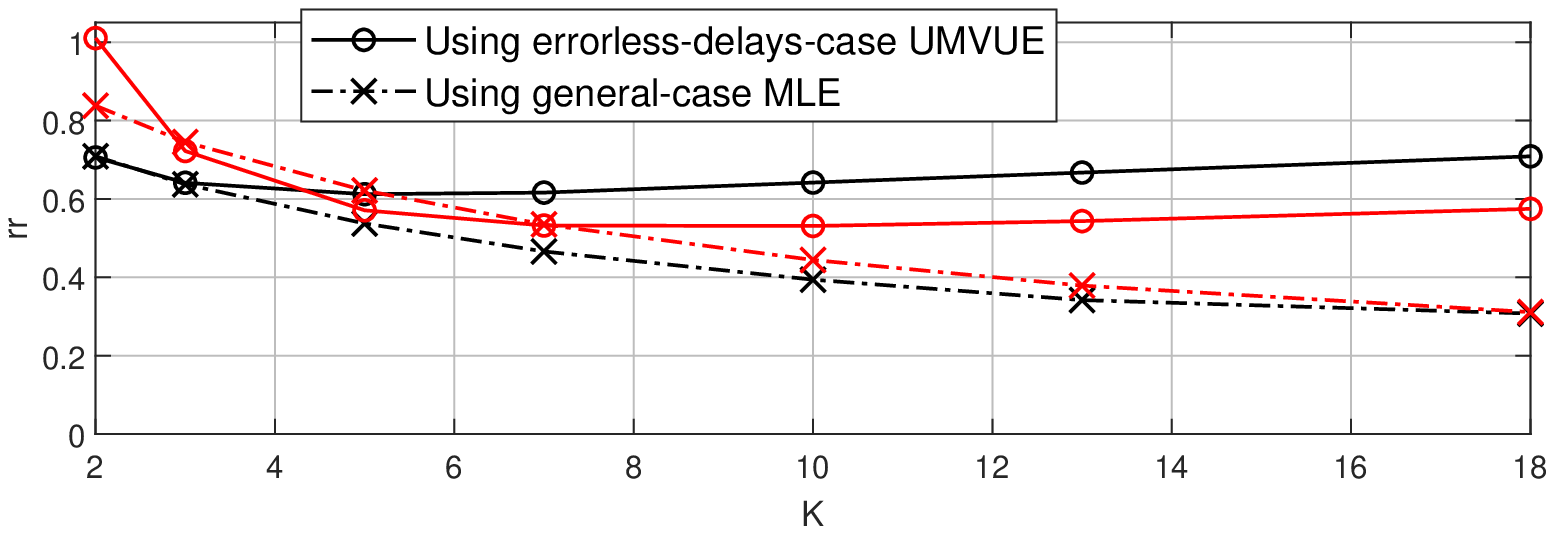} 
    \label{fig:evalOverKCsod500}}} \ \\
  \subfloat[Impact of $\sigma$; number of common det. MPCs fixed to $K=18$]{\shortstack[l]{
		\psfrag{csod}{\hspace{-24mm}\raisebox{-.2mm}{\scriptsize normalized standard deviation $\cd\sigma/d$ of delay error}}
		\psfrag{rb}{\hspace{-5.1mm}\raisebox{0mm}{\scriptsize relative bias}}
		\includegraphics[width=\columnwidth,trim=4mm  0mm 10mm 3mm,clip=true]{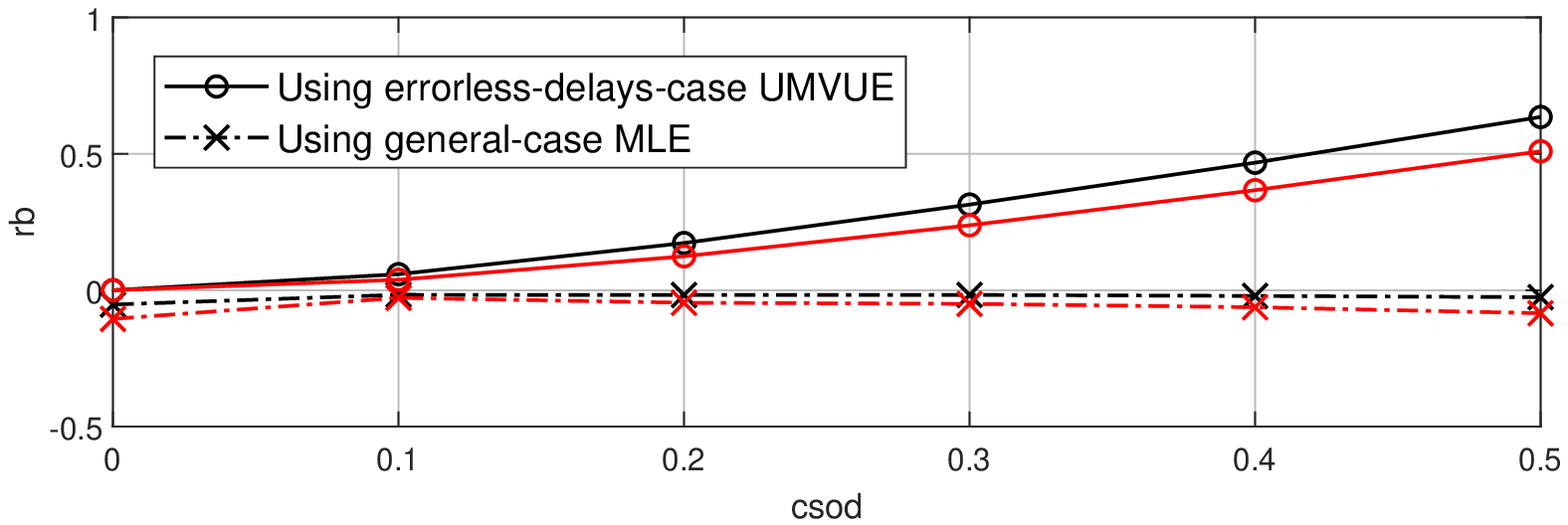} \\
    \psfrag{csod}{\hspace{-24mm}\raisebox{-.2mm}{\scriptsize normalized standard deviation $\cd\sigma/d$ of delay error}}
		\psfrag{rr}{\hspace{-7.5mm}\raisebox{0mm}{\scriptsize relative RMSE}}
		\includegraphics[width=\columnwidth,trim=4mm -3mm 10mm 3mm,clip=true]{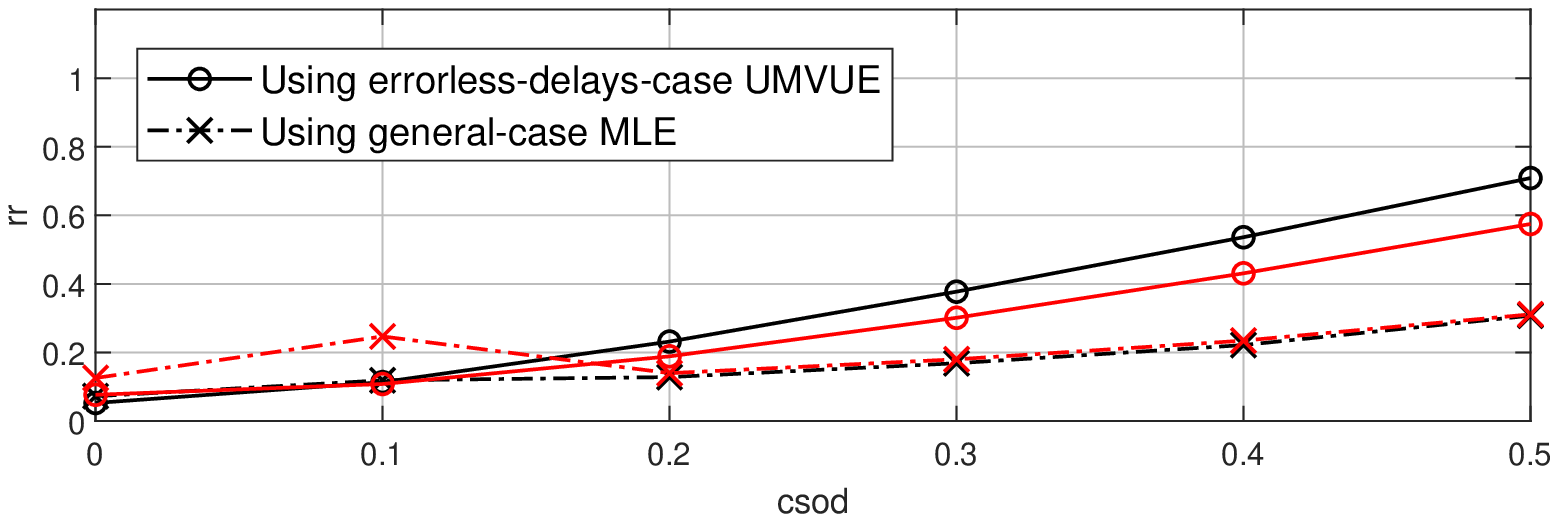}
    \label{fig:evalOverCsodK18}}}
    \caption{The plots show the dependence of the relative bias $\mathrm{E}[\dh\,] / d - 1$ and the relative RMSE $\mathrm{std}[\dh\,] / d$ on the number of common detected delays $K$ and the standard deviation $\sigma$ of the Gaussian error on each $\S_k$. The black and red graphs represent synchronous and asynchronous estimates, respectively.}
    \label{fig:eval}
\end{figure}

The errors $\err_k$ on the delay differences, which stem from delay extraction errors, cause additional performance degradation. We will now evaluate how the relative error $(\hat{d}-d)/d$ of various distance estimates is affected by independent Gaussian errors
$\err_k \sim \mathcal{N}(0,\sigma^2)$ while the statistics of $\S_k$ are according to the assumptions in Section~\ref{sec:estimators}. We do not assume a specific setup geometry but instead specify the ratio of $\cd\sigma$ (the distance-translated error standard deviation) to $d$ as it determines the statistics of the relative error.
To implement the general-case MLEs we use \eqref{eq:softIndGaussian} in \eqref{eq:dMLSyncNoisy} and \eqref{eq:dMLAsynNoisy} and solve the respective optimization problems with an iterative Gauss-Newton algorithm. Fig.~\ref{fig:eval} shows the impact of $\sigma$ and $K$ on the relative error. 

When $\cd\sigma/d$ is considerably large, we observe that the estimators designed for errorless delay extraction are heavily distorted. This effect is even amplified with increasing $K$ (giving rise to more outliers). In this case, the asynchronous estimate outperforms the synchronous estimate as it uses not one but two delay-differences (the extrema) which amounts to some error averaging.

In the high $\cd\sigma/d$ regime, the general-case estimators perform much better because they are tailored to the observation statistics at hand. We observe that bias and RMSE converge to zero with increasing $K$ and that the bias is very small even at high error levels. If $\cd\sigma$ is very small, e.g. with a capable ultra-wideband system, then $\cd\sigma/d$ is significant only for small $d$. This short-range regime is of particular practical interest though, as argued in Sec.~\ref{sec:intro}.

We conclude that the simple proposed estimates \eqref{eq:dUBSync} and \eqref{eq:dUBAsyn} perform well if $\cd\sigma/d$ is less than about $0.1$. For larger error levels the general-case estimates \eqref{eq:dMLSyncNoisy} and \eqref{eq:dMLAsynNoisy} should instead be used, e.g. at close proximity or with small bandwidth.

\newcommand{\mySF}[3]{\centering\subfloat[#2]{
\psfrag{xm}{\hspace{1.6mm}\raisebox{0.3mm}{\scriptsize $x\ \mathrm{[m]}$}}
\psfrag{ym}{\hspace{0.0mm}\raisebox{0.5mm}{\scriptsize $y\ \mathrm{[m]}$}}
\includegraphics[width=.98\columnwidth,trim={11mm 23mm -6mm 28mm},clip=true]{#1}\label{#3}}
\put(-118,34){\footnotesize{node A (static)}}
\put(-218.7,50){\footnotesize{\textcolor[rgb]{1,1,1}{\begin{tabular}{c}observer\\node\end{tabular}}}}
\put(-10,25){\rotatebox{90}{\footnotesize distance est. error $\mathrm{[m]}$}}}

\newcommand{\mySO}[3]{\centering\subfloat[#2]{
\hspace{.7mm}
\includegraphics[width=.97\columnwidth,trim={11mm 35mm -6mm 28mm},clip=true]{#1}\label{#3}}
\put(-10,5){\rotatebox{90}{\footnotesize distance est. error $\mathrm{[m]}$}}}

\begin{figure}[!b]
\vspace{-4mm}
\mySF{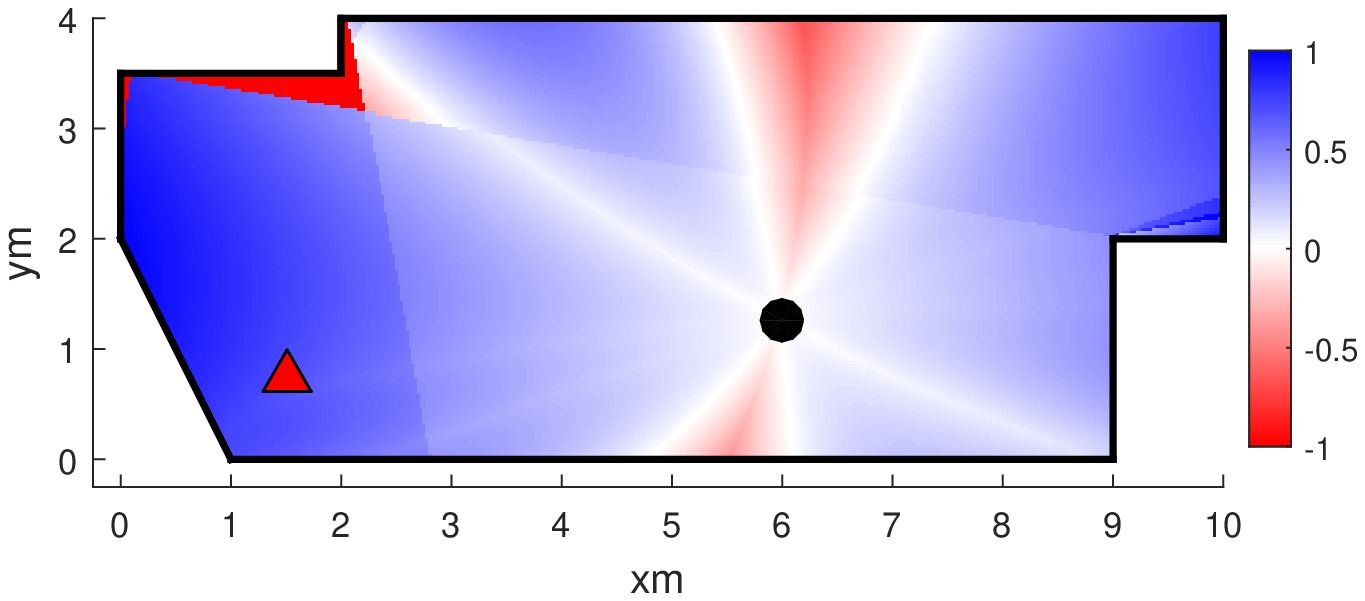}{Synchronous case, single observer}{fig:RTSync}  \ \\[1.5mm]
\mySO{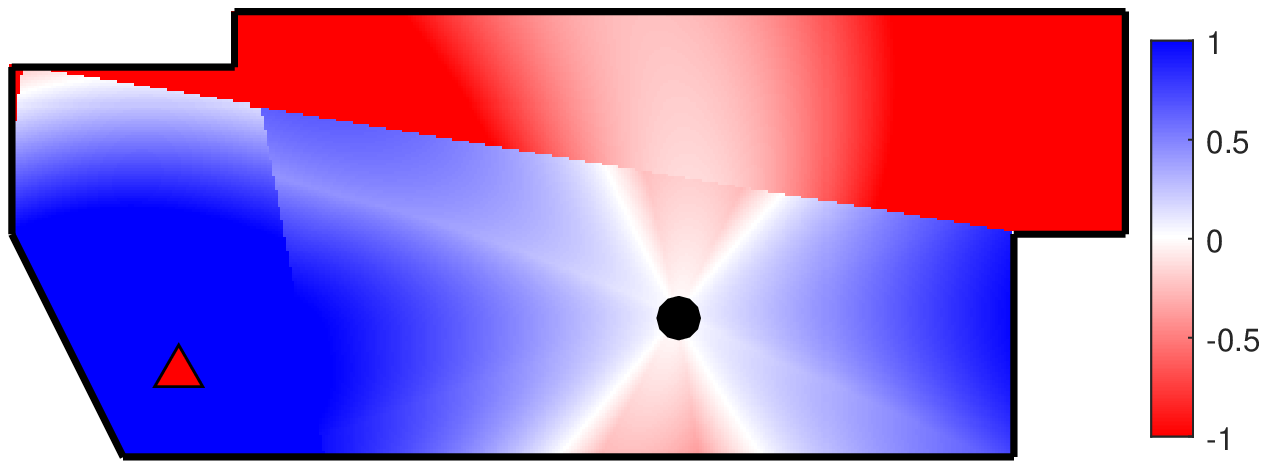}{Asynchronous case, single observer}{fig:RTAsyn} \ \\[1.5mm]
\mySO{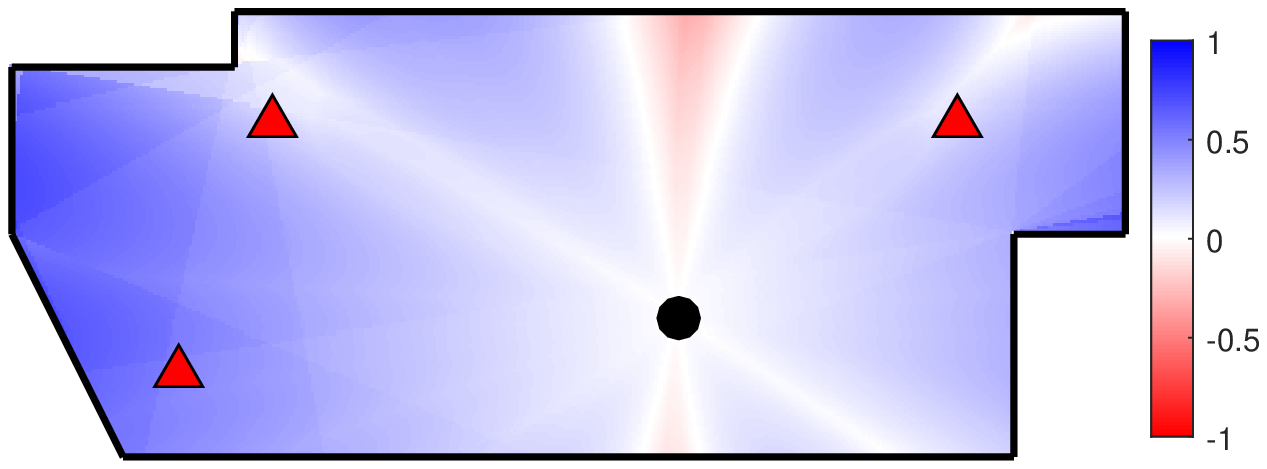}{Asynchronous case, three observers}{fig:RTAsynMulti}
\caption{Distance estimation error between a static node A ($\bullet$) and a mobile node B (anywhere in the room) in an indoor environment of the shown floor plan. This experiment assumes errorless delay extraction (and uses the according UMVUEs), yet estimation errors occur because of the reliance on unknown MPC directions. We use an MPC detection threshold $\mathrm{SINR} \geq 0\,\mathrm{dB}$ assuming diffuse multipath and additive noise at $1\,\mathrm{GHz}$ bandwidth.}
\label{fig:RT}
\end{figure}

\subsection{Impact of realistic MPC directions}
\label{sec:RealisticSim}

We will now evaluate how the proposed estimates, which were designed for the propagation assumptions of Sec.~\ref{sec:estimators}, perform in realistic indoor propagation conditions and with a signaling bandwidth of $1\,\mathrm{GHz}$ for estimation of the CIRs. In particular, we consider a room with the floor plan shown in Fig.~\ref{fig:RTSync} and a static observer node, a static node A, and a mobile node B that can be located anywhere in the room. We employ ray tracing to simulate reflection paths of up to three bounces whereby each bounce is assumed to cause $3\,\mathrm{dB}$ attenuation \cite{WitrisalSPM2016,LeitingerJSAC2015}. We consider reflections via the side walls as well as the floor and ceiling. The assumed room height is $3\,\mathrm{m}$ and all devices are $1.2\,\mathrm{m}$ above the floor.

To obtain practically meaningful results we need to define a criterion for the detection of a MPC. We use the detection threshold $\mathrm{SINR}_k \geq 0\,\mathrm{dB}$ based on the signal-to-interference-plus-noise ratio whereby the interference is due to diffuse multipath propagation. In particular, we employ the definition
$\mathrm{SINR}_k = |a_k|^2 / (N_0 + T_p S_\nu(\tau_k))$
from \cite[Eq. 14]{LeitingerJSAC2015} where $a_k$ is the $k$-th path amplitude (which is subject to free-space path loss), $N_0$ the single-sided noise spectral density, $T_p$ the effective pulse duration (inversely proportional to bandwidth), and $S_\nu(\tau)$ the power delay profile of the diffuse multipath portion in the CIRs. Following the proposal of \cite[Tab. 1]{LeitingerJSAC2015}, we choose a double-exponential $S_\nu(\tau)$ with $5\,\mathrm{ns}$ rise time, $20\,\mathrm{ns}$ decay time constant, and $1.16 \cdot 10^{-6}$ normalized power.

Fig.~\ref{fig:RT} shows the distance estimation error (the color at any point $x,y$ in the room marks the error when the mobile node B is at that position) for the UMVUEs in the absence of extraction errors. We observe a significant performance advantage with synchronization in Fig.~\ref{fig:RTSync} over the asynchronous case in Fig.~\ref{fig:RTAsyn}. The reason is that with a single observer, $K$ is small and the MPC directions tend to be similar to the LOS direction rather than uniformly distributed. This heavily impairs the asynchronous estimate.
With the three observer deployment of Fig.~\ref{fig:RTAsynMulti} however, $K$ increases vastly (from $6$ or $7$ to about $20$ for most positions) and the $\e_k$ are spread more evenly, which results in great performance even in asynchronous mode.

\begin{figure}[!ht]
    \centering
    \psfrag{d}{\hspace{-7.2mm}\raisebox{-.2mm}{\scriptsize distance $d\ \mathrm{[m]}$}}
		\psfrag{r}{\hspace{-9.8mm}\raisebox{0mm}{\scriptsize distance RMSE $\mathrm{[m]}$}}
    \includegraphics[width=\columnwidth,trim={11mm 0mm 16mm 0mm},clip=true]{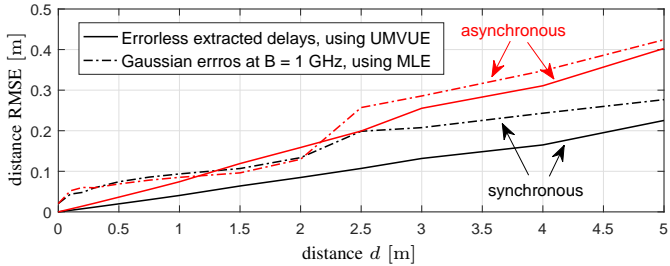}
    \caption{For the three-observer setup of Fig.~\ref{fig:RTAsynMulti}, the plot shows the RMSE of distance estimation as a function of distance $d$ around the static node. The plot compares distance estimation performance under errorless and erroneous delay extraction.  We assume $1\,\mathrm{GHz}$ bandwidth and use the MPC detection criterion $\mathrm{SINR}_k \geq 0\,\mathrm{dB}$. The values of the near-constant slopes conform with the predictions of Sec.~\ref{sec:EvalUnknown} (the RMSE $\propto d$ for errorless extraction).}
    \label{fig:Circles}
    \vspace{-1.7mm}
\end{figure}

For the three-observer setup, Fig.~\ref{fig:Circles} shows the RMSE as a function of $d$ around the static node (computed from error realizations on a circle of radius $d$). We observe almost constant slopes, consistent with the scaling behavior described in Sec.~\ref{eq:EvalExtract}. At $d = 3\,\mathrm{m}$, we measure a relative RMSE of $4.39\%$ (synchronous) and $8.51\%$ (asynchronous) which compare to analytical projections of $4.77\%$ and $7.10\%$, respectively, from \eqref{eq:stdUBSync} and \eqref{eq:stdUBAsyn}. We infer that this setup faces no performance degradation due to non-uniform $\e_k$ with established synchronization and just a slight degradation in the asynchronous case. 

Fig.~\ref{fig:Circles} also shows the performance for erroneous MPC extraction and using the general-case MLEs. The chosen error model is $\err_k \sim \mathcal{N}(0,\sigma_k^2)$ where $\sigma_k^2 = \sigma\Nk{A}^2 + \sigma\Nk{B}^2$ is the sum of the variances of assumed independent Gaussian errors on $\tau\Nk{A}$ and $\tau\Nk{B}$, respectively, which are set to the minimum variance according to the Cram\'er-Rao lower bound (CRLB) for delay extraction (neglecting path overlap) in diffuse multipath and noise as presented in \cite{LeitingerJSAC2015}. For the details we refer to \cite[Sec. III.-B]{LeitingerJSAC2015}. The resulting $\cd\sigma_k$ are between $5\,\mathrm{cm}$ and $10\,\mathrm{cm}$ and thus the extraction errors should have a significant impact for about $d \leq 1\,\mathrm{m}$ (i.e. $\cd\sigma_k/d \leq 0.1$), which agrees with the numerical results in large part. Extraction errors obviously impair the performance but do not change the order of magnitude of the estimation errors, which stay below $20\,\mathrm{cm}$ in a circle of at least $d \leq 2\,\mathrm{m}$ around the static node. We conclude that the proposed distance estimators are viable in realistic conditions.

\section{Technological Comparison and Opportunities}\label{sec:sota}
A performance comparison to related distance estimation schemes is in order. With the parameters of our evaluation in Fig.~\ref{fig:Circles}, a TOA distance estimate to fixed infrastructure (e.g. from node A to an observer position) would have an RMSE lower bound \cite{LeitingerJSAC2015} of $\cd\sigma_{\text{A},1} \approx 2.7\,\mathrm{cm}$, which implies high accuracy but requires LOS and perfect time synchronization. When a synchronization error $\epsilon$ occurs, a TOA estimate suffers a distance error $\cd\epsilon$ (e.g. $\cd\epsilon = 30\,\mathrm{cm}$ error from just $\epsilon = 1\,\mathrm{ns}$) which is particularly severe for short distances. Yet, to the best of our knowledge, distributed synchronization with sub-$\mathrm{ns}$ precision is currently not feasible with reasonable complexity.
Our scheme compares well to reported TOA ranging errors of up to $2\,\mathrm{m}$ in \cite{Chen2017} ($500\,\mathrm{MHz}$ bandwidth) and \cite{AlaviCL2006} ($1\,\mathrm{GHz}$) or up to $10\,\mathrm{m}$ in \cite{LiJSAC2015} ($125\,\mathrm{MHz}$) and \cite{AlaviCL2006} ($200\,\mathrm{MHz}$), although a thorough comparison is out of scope.

As indicated in Sec.~\ref{sec:intro}, our proposal has various promising applications in indoor localization. Due to the conceptual individuality, a direct performance comparison to existing schemes is not possible at this point. Instead we highlight the major technological opportunities and benefits in the following.

The absence of synchronization and LOS requirements qualifies the proposal for localization in dense and crowded settings.
It allows for accurate ranging between low-complexity nodes, which only need to transmit pilot sequences and do not require high-resolution wideband receivers (only the observers do).
The scheme is thus a prime candidate for estimating distances between mobiles for the purpose of network localization and, at that, does not require interaction of the mobiles. Thereby, the fact that mobiles can be observers (as knowledge of observer positions is not required) promises particularly great performance scaling with network density: $N-2$ out of $N$ mobiles can be observers for each distance estimation.


As described in Sec.~\ref{sec:intro}, the proposal can be used for localization via trilateration when distances to multiple stationary nodes (henceforth called beacons) are obtained. The low complexity requirements allow for battery-powered transmit-only beacons without a wired connection. They can thus be deployed  easily and in vast numbers. This is a major advantage over state-of-the-art systems, e.g., TDOA systems which require precisely synchronized anchor infrastructure with LOS coverage. A vast amount of distance estimates between many beacons and mobiles together with all the inter-mobile distances promises accurate and robust network localization. The mobiles can also be of low complexity, as all processing and hardware complexity could be pushed to fixed observer infrastructure.

The single-beacon ranging application of Fig.~\ref{fig:AppBeacon} is similar to wideband location fingerprinting \cite{SteinerTSP2010} but does not rely on offline training: The beacon-to-observer CIR can be estimated online, enabling robust operation in dynamic environments. 

Our proposal utilizes multipath propagation without using any specific knowledge about the environment. This is in contrary to multipath-assisted localization  \cite{WitrisalSPM2016,LeitingerJSAC2015} which uses an a-priori known floor plan or online learning in order to utilize reflected paths for localization (which softens requirements on LOS conditions and number of anchors).

It is noteworthy that delay differences have been employed for microphone synchronization in audio engineering \cite{Ono2009}.

\section{Summary \& Outlook}\label{sec:summary}
We proposed a novel paradigm to estimate the distance between two wireless nodes by a comparison of the impulse responses of their channels to auxiliary observer nodes.
Based on the multipath delay structure of the CIRs, we derived distance estimators and their properties for different relevant cases.
A numerical evaluation showed that an accuracy of $20\,\mathrm{cm}$ can be achieved over large parts of a typical-size office room when using three observers (which could be other mobiles), $1\,\mathrm{GHz}$ signaling bandwidth, and no synchronization requirements whatsoever.
We described how this scheme could improve indoor localization in various use cases: (i) spacious buildings because the distributed infrastructure can be mostly simple transmit-only beacons, (ii) crowded settings because it does not rely on LOS connection, or (iii) network localization because it promises to be well-suited for the estimation of small distances between mobiles. As the next steps towards an implementation, future work should evaluate the performance of the scheme with practical MPC extraction and association techniques and the impact of clock drift of different observers in the time lag between the estimation of the two channels.

\appendices
\section*{Appendix}
This appendix outlines the derivations of the analytic results in Sec.~\ref{sec:estimators} and Sec.~\ref{sec:EvalUnknown}.
\addtolength{\topmargin}{.03in}
\subsection{Delays extracted without error; synchronous clocks}
To estimate $d$ from the i.i.d. samples
$\cd\S_k \sim \mathcal{U}(-d,d)$,
we can equivalently consider the i.i.d. $\cd|\S_k| \sim \mathcal{U}(0,d)$. 
The MLE \eqref{eq:dMLSync} is easily found by maximizing the conditional PDF.

To analyze the statistics of the estimates, consider the i.i.d. $x_k = \f{c}{d}\,|\S_k| \sim \mathcal{U}(0,1)$. We employ the order statistics \cite{David1970} 
of $x_k$ via indexation $x_{(k)}$ such that
$x_{(1)} \leq x_{(2)} \leq \ldots \leq x_{(K)}$.
The key consequence is 
$x_{(k)} \sim \mathrm{Beta}(k,K-k+1)$
and thus
\begin{align}
& \mathrm{E}[x_{(k)}] = \f{k}{K+1} \ , &
& \mathrm{var}[x_{(k)}] = \f{k(K-k+1)}{(K+1)^2(K+2)} \ .
\label{eq:OrderMeanVar}
\end{align}
Now
$\mathrm{E}[\dh\synNL] = d\,\mathrm{E}[x_{(K)}]$ and
$\mathrm{std}[\dh\synNL] = d\,\mathrm{var}[x_{(K)}]^{1/2}$
with \eqref{eq:OrderMeanVar} yield the remaining results; also that \eqref{eq:dUBSync} is unbiased. It is thus the UMVUE by the Lehmann–Scheffé theorem as $\max \big\{ |\S_k| \big\}$ is a complete sufficient statistic.

\subsection{Delays extracted without error; asynchronous clocks}

From the i.i.d. samples
$\cd\SA_k \sim \mathcal{U}(\cd\epsilon-d,\cd\epsilon+d)$,
the joint MLE \eqref{eq:dMLAsyn}, \eqref{eq:eMLAsyn} of $d$ and $\epsilon$ is found by careful maximization of the conditional PDF.
To prove the bias and RMSE results, consider the i.i.d.
$\tilde{x}_k = \f{1}{2} \left( \f{c}{d} \S_k + 1 \right) \sim \mathcal{U}(0,1)$ and their order statistics $\tilde{x}_{(k)}$ with mean and variance in \eqref{eq:OrderMeanVar}. With $\SA_k = \S_k + \epsilon$, we find
$\max \SA_k - \min \SA_k = \max \S_k - \min \S_k = \f{2d}{c} (\tilde{x}_{(K)} - \tilde{x}_{(1)})$
and further
$\mathrm{E}[\dh\asyNL] = d\left( \mathrm{E}[\tilde{x}_{(K)}] - \mathrm{E}[\tilde{x}_{(1)}] \right) = d\,\f{K-1}{K+1}$. Thus $\dh\UB\asyNL$ is unbiased. It is the UMVUE because it uses the minimum and maximum sample which form a complete sufficient statistic of the uniform distribution.
%
%
The RMSE follows from
\begin{multline}
\mathrm{var}\big[ \dh\asyNL \big]
= \mathrm{var}\big[ d\,\big( \tilde{x}_{(K)} - \tilde{x}_{(1)} \big)\big] \\
= d^2 \!\left( \mathrm{var}[\tilde{x}_{(K)}] + \mathrm{var}[\tilde{x}_{(1)}] - 2\,\mathrm{cov}[\tilde{x}_{(K)},\tilde{x}_{(1)}] \right).
\end{multline}
We argue $\mathrm{cov}[\tilde{x}_{(K)},\tilde{x}_{(1)}] \approx 0$ for sufficiently large $K$ and through \eqref{eq:OrderMeanVar} obtain
$\mathrm{var}[ \dh\asyNL ] \approx d^2 \f{2K}{(K+1)^2(K+2)}$
and finally \eqref{eq:stdUBAsyn} by expanding
$\mathrm{std}[ \dh\UB\asyNL ] = \f{K+1}{K-1} \mathrm{var}[ \dh\asyNL ]^{1/2}$. For $\eh\UB\asyNL$ the RMSE and zero bias follow analogously.

\subsection{General case with synchronous clocks}

The likelihood function (LHF) of $d$ from one observation $\SN_k = \S_k + \err_k$ is the conditional PDF given by the convolution
\begin{align}
f(\SN_k\,|\,d) 
= \int_\mathbb{R} f_{\err_k}(\err) f_{\S_k | d}(\SN_k - \err \,|\, d) d\err .
\end{align}
With $\S_k | d \sim \mathcal{U}(-d/c,d/c)$ from \eqref{eq:UniformShift} we furthermore obtain
\begin{equation}
f(\SN_k|d) 
= \f{c}{2d} \int_{\SN_k-d/c}^{\SN_k+d/c} f_{\err_k}(\err) d\err
= \f{c}{2d}\,I_k(\SN_k,d)
\label{eq:SingleObsLHF}
\end{equation}
where we use definition \eqref{eq:softInd}. The LHF of $d$ given $\SN_1, \ldots, \SN_K$ is the product of the individual $f(\SN_k|d)$, i.e.
\vspace{-1mm}
\begin{align}
L(d) = f(\SN_1,\ldots,\SN_K \,|\, d) = \left( \f{c}{2d} \right)^K \prod_{k=1}^K \, I_k(\SN_k,d)
\label{eq:LhfNoisySync}
\end{align}
\ \\[-2.5mm]
because the observations are assumed statistically independent. A distance $d$ that maximizes $L(d)$ is an MLE, giving \eqref{eq:dMLAsynNoisy}.

\subsection{General case with asynchronous clocks}

The only difference to the above case are the asynchronous $\SNA_k = \SN_k + \epsilon$. As $\epsilon$ is modeled non-random and $\SN_k = \SNA_k - \epsilon$,
\begin{equation}
f_{\SNA_k | d, \epsilon}(\SNA_k | d, \epsilon)
= f_{\SN_k | d}(\SNA_k - \epsilon | d) 
= \f{c}{2d}\,I_k(\SNA_k - \epsilon,d).
\end{equation}
The joint MLE \eqref{eq:dMLAsynNoisy} given $\SN_1, \ldots, \SN_K$ is due to the LHF
\vspace{-1mm}
\begin{align}
\tilde{L}(d,\epsilon) = \left( \f{c}{2d} \right)^K \prod_{k=1}^K \, I_k(\SNA_k - \epsilon,d).
\label{eq:LhfNoisyAsyn}
\end{align}


\section*{Acknowledgment}

We would like to thank Malte G\"oller for initial investigations on the subject in his master's thesis 
as well as Klaus Witrisal and his group at Graz University of Technology for providing the ray tracer used in \cite{LeitingerJSAC2015,WitrisalSPM2016}.

\IEEEtriggeratref{0}
\bibliographystyle{IEEEtran}
\bibliography{../../../GD}

\begin{thebibliography}{10}
\providecommand{\url}[1]{#1}
\csname url@samestyle\endcsname
\providecommand{\newblock}{\relax}
\providecommand{\bibinfo}[2]{#2}
\providecommand{\BIBentrySTDinterwordspacing}{\spaceskip=0pt\relax}
\providecommand{\BIBentryALTinterwordstretchfactor}{4}
\providecommand{\BIBentryALTinterwordspacing}{\spaceskip=\fontdimen2\font plus
\BIBentryALTinterwordstretchfactor\fontdimen3\font minus
  \fontdimen4\font\relax}
\providecommand{\BIBforeignlanguage}[2]{{%
\expandafter\ifx\csname l@#1\endcsname\relax
\typeout{** WARNING: IEEEtran.bst: No hyphenation pattern has been}%
\typeout{** loaded for the language `#1'. Using the pattern for}%
\typeout{** the default language instead.}%
\else
\language=\csname l@#1\endcsname
\fi
#2}}
\providecommand{\BIBdecl}{\relax}
\BIBdecl

\bibitem{BuehrerPIEEE2018}
R.~M. Buehrer, H.~Wymeersch, and R.~M. Vaghefi, ``Collaborative sensor network
  localization: Algorithms and practical issues,'' \emph{Proceedings of the
  IEEE}, vol. 106, no.~6, pp. 1089--1114, 2018.

\bibitem{LiJSAC2015}
S.~Li, M.~Hedley, and I.~B. Collings, ``New efficient indoor cooperative
  localization algorithm with empirical ranging error model,'' \emph{IEEE
  Journal on Selected Areas in Communications}, vol.~33, no.~7, pp. 1407--1417,
  2015.

\bibitem{LiuTSP2018}
Y.~Liu, Y.~Shen, D.~Guo, and M.~Z. Win, ``Network localization and
  synchronization using full-duplex radios,'' \emph{IEEE Transactions on Signal
  Processing}, vol.~66, no.~3, pp. 714--728, 2018.

\bibitem{MazuelasTSP2018}
S.~Mazuelas, A.~Conti, J.~C. Allen, and M.~Z. Win, ``Soft range information for
  network localization,'' \emph{IEEE Transactions on Signal Processing},
  vol.~66, no.~12, pp. 3155--3168, 2018.

\bibitem{SchultenVTC2019}
H.~Schulten, M.~Kuhn, R.~Heyn, G.~Dumphart, A.~Wittneben, and F.~Tr\"osch, ``On
  the crucial impact of antennas and diversity on {BLE RSSI}-based indoor
  localization,'' in \emph{IEEE Vehicular Technology Conference (VTC Spring)},
  May 2019.

\bibitem{DardariPIEEE2009}
D.~Dardari, A.~Conti, U.~Ferner, A.~Giorgetti, and M.~Z. Win, ``Ranging with
  ultrawide bandwidth signals in multipath environments,'' \emph{Proceedings of
  the IEEE}, vol.~97, no.~2, pp. 404--426, 2009.

\bibitem{WymeerschPIEEE2009}
H.~Wymeersch, J.~Lien, and M.~Z. Win, ``Cooperative localization in wireless
  networks,'' \emph{Proceedings of the IEEE}, vol.~97, no.~2, pp. 427--450,
  2009.

\bibitem{AlaviCL2006}
B.~Alavi and K.~Pahlavan, ``Modeling of the {TOA}-based distance measurement
  error using {UWB} indoor radio measurements,'' \emph{IEEE communications
  letters}, vol.~10, no.~4, pp. 275--277, 2006.

\bibitem{JourdanTAES2008}
D.~B. Jourdan, D.~Dardari, and M.~Z. Win, ``Position error bound for {UWB}
  localization in dense cluttered environments,'' \emph{IEEE transactions on
  aerospace and electronic systems}, vol.~44, no.~2, 2008.

\bibitem{WitrisalSPM2016}
K.~Witrisal, P.~Meissner, E.~Leitinger, Y.~Shen, C.~Gustafson, F.~Tufvesson,
  K.~Haneda, D.~Dardari, A.~F. Molisch, A.~Conti \emph{et~al.}, ``High-accuracy
  localization for assisted living: {5G} systems will turn multipath channels
  from foe to friend,'' \emph{IEEE Signal Processing Magazine}, vol.~33, no.~2,
  pp. 59--70, 2016.

\bibitem{Gani2016}
M.~O. Gani, G.~M.~T. Ahsan, D.~Do, W.~Drew, M.~Balfas, S.~I. Ahamed, M.~Arif,
  and A.~J. Kattan, ``An approach to localization in crowded area,'' in
  \emph{e-Health Networking, Applications and Services (Healthcom), 2016 IEEE
  18th International Conference on}, 2016.

\bibitem{MolischPIEEE2009}
A.~F. Molisch, ``Ultra-wide-band propagation channels,'' \emph{Proceedings of
  the IEEE}, vol.~97, no.~2, pp. 353--371, 2009.

\bibitem{DokmanicPNAS2013}
I.~Dokmani{\'c}, R.~Parhizkar, A.~Walther, Y.~M. Lu, and M.~Vetterli,
  ``Acoustic echoes reveal room shape,'' \emph{Proceedings of the National
  Academy of Sciences}, vol. 110, no.~30, pp. 12\,186--12\,191, 2013.

\bibitem{LeitingerJSAC2015}
E.~Leitinger, P.~Meissner, C.~R{\"u}disser, G.~Dumphart, and K.~Witrisal,
  ``Evaluation of position-related information in multipath components for
  indoor positioning,'' \emph{IEEE Journal on Selected Areas in
  Communications}, vol.~33, no.~11, pp. 2313--2328, 2015.

\bibitem{DumphartPIMRC2016}
G.~Dumphart and A.~Wittneben, ``Stochastic misalignment model for
  magneto-inductive {SISO} and {MIMO} links,'' in \emph{IEEE International
  Symposium on Personal, Indoor, and Mobile Radio Communications (PIMRC)}, Sep.
  2016.

\bibitem{Qi2003}
Y.~Qi, ``Wireless geolocation in a non-line-of-sight environment,'' Ph.D.
  dissertation, Princeton University, 2003.

\bibitem{Chen2017}
Y.~Chen, ``Evaluating off-the-shelf hardware for indoor positioning,'' Master's
  thesis, Lund University, 2017.

\bibitem{SteinerTSP2010}
C.~Steiner and A.~Wittneben, ``Low complexity location fingerprinting with
  generalized {UWB} energy detection receivers,'' \emph{IEEE Transactions on
  Signal Processing}, vol.~58, no.~3, pp. 1756--1767, 2010.

\bibitem{Ono2009}
N.~Ono, H.~Kohno, N.~Ito, and S.~Sagayama, ``Blind alignment of asynchronously
  recorded signals for distributed microphone array,'' in \emph{Applications of
  Signal Processing to Audio and Acoustics (WASPAA), IEEE Workshop on}, 2009,
  pp. 161--164.

\bibitem{David1970}
H.~A. David and H.~N. Nagaraja, \emph{Order statistics}.\hskip 1em plus 0.5em
  minus 0.4em\relax Wiley, 1970.

\end{thebibliography}



\clearpage

\begin{figure*}[!ht]
The following material is supplementary.\hfill 

\centering
\subfloat[Likelihood function \eqref{eq:LhfNoisyAsyn} and associated estimates for delay differences without extraction errors, i.e. for $\err_k \equiv 0$ and $I_k(\SN_k,d) = \IndFunc_{[-d,d\,]}(\cd \SN_k)$]{
    \centering
    \psfrag{ddm1}{\raisebox{-.5mm}{\hspace{-2.5mm}\scriptsize distance $d\ \mathrm{[m]}$}}
    \psfrag{coens1}{\raisebox{.5mm}{\hspace{-5.6mm}\scriptsize clock offset $\epsilon\ \mathrm{[ns]}$}}
    \psfrag{LH1}{\raisebox{-1mm}{\hspace{-3.8mm}\scriptsize Likelihood}}
    \includegraphics[width=1.2\columnwidth,trim=18mm 87mm 18mm 8mm,clip=true]{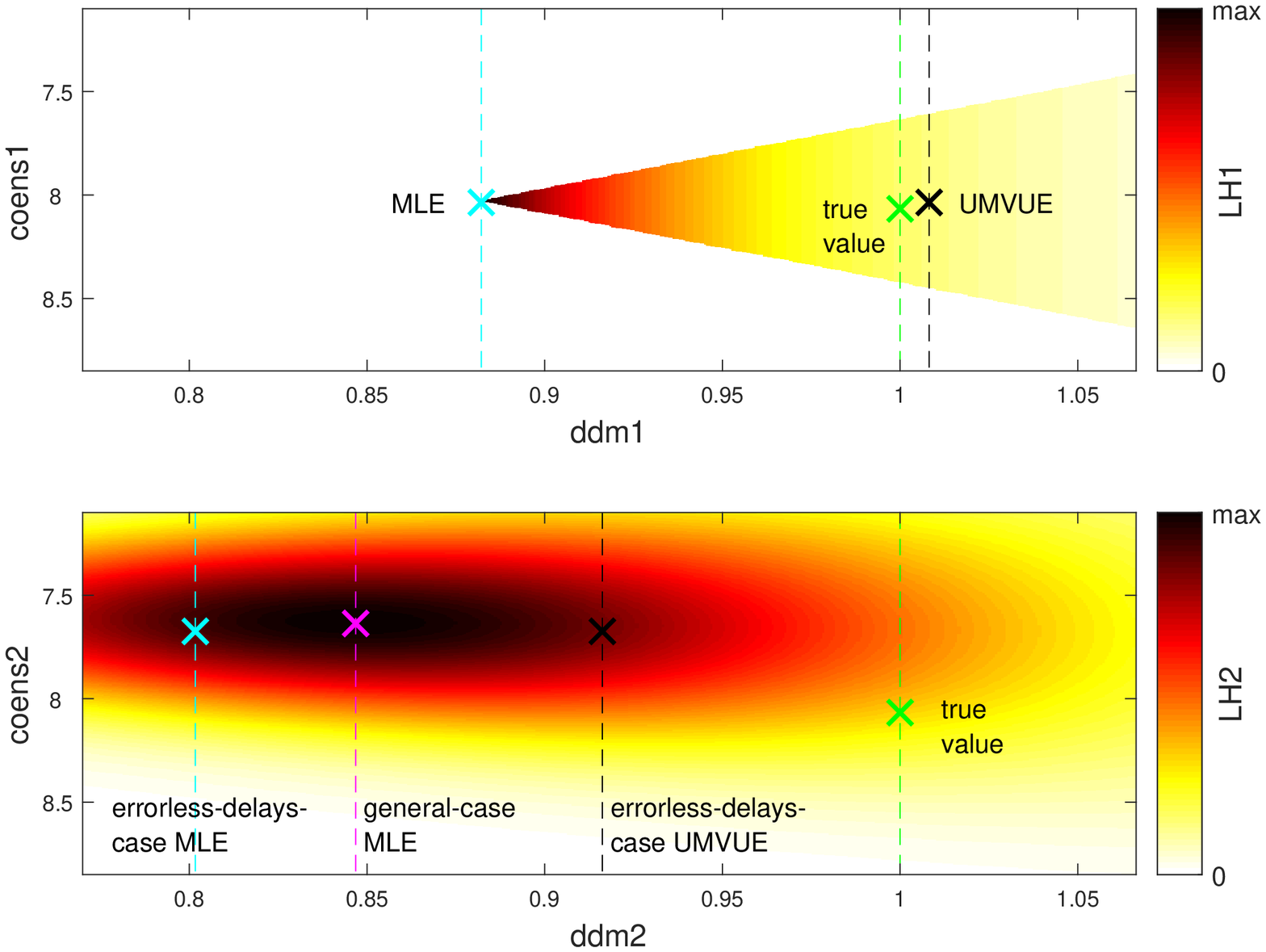}
    \label{fig:lhfErrorless}}
  \ \\[1mm]
  \subfloat[Likelihood function \eqref{eq:LhfNoisyAsyn} and various estimates for delay differences with Gaussian errors $\err_k$, i.e. using \eqref{eq:softIndGaussian}]{
    \centering
    \psfrag{ddm2}{\raisebox{-.5mm}{\hspace{-2.5mm}\scriptsize distance $d\ \mathrm{[m]}$}}
    \psfrag{coens2}{\raisebox{.5mm}{\hspace{-5.6mm}\scriptsize clock offset $\epsilon\ \mathrm{[ns]}$}}
    \psfrag{LH2}{\raisebox{-1mm}{\hspace{-3.8mm}\scriptsize Likelihood}}
    \includegraphics[width=1.2\columnwidth,trim=18mm 6mm 18mm 90mm,clip=true]{TauShiftAsyncLhfColor}
    \label{fig:lhfGauss}}
\caption{Color plots of examples for the asynchronous-case likelihood function $\tilde{L}(d,\epsilon)$. Plot (a) is based on observations $\SA_k$ without extraction errors and (b) is based on $\SNA_k \sim \mathcal{N}(\SA_k,\sigma^2)$. We assumed a true distance $d = 1\,\mathrm{m}$, an error level of $\cd\sigma/d = 0.2$, and a total of $K = 15$ MPCs.
}
\label{fig:TauShiftAsyncLhfColorPlot}
\end{figure*}

\begin{figure*}[!b]
\centering\subfloat[Single observer]{
\psfrag{xm}{\hspace{1.6mm}\raisebox{0.3mm}{\scriptsize $x\ \mathrm{[m]}$}}
\psfrag{ym}{\hspace{0.0mm}\raisebox{0.3mm}{\scriptsize $y\ \mathrm{[m]}$}}
\includegraphics[width=1.2\columnwidth,trim={9mm 25mm 7mm 28mm},clip=true]{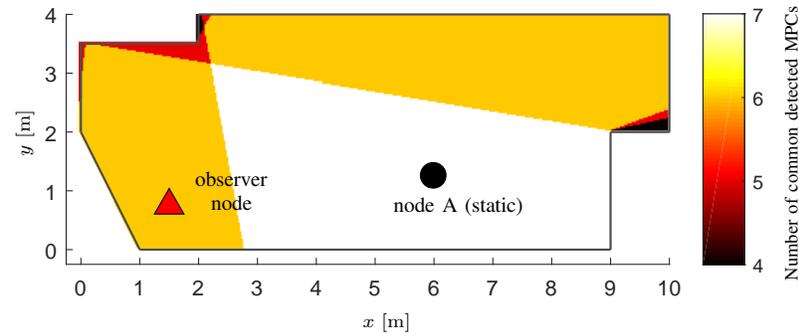}\label{fig:NoMPC}}
\put(-161,47){\footnotesize{node A (static)}}
\put(-242,52){\footnotesize{\begin{tabular}{c}observer\\node\end{tabular}}}
\put(-13,21){\rotatebox{90}{\scriptsize Number of common detected MPCs}}

\centering\subfloat[Three observers]{
\psfrag{xm}{\hspace{1.6mm}\raisebox{0.3mm}{\scriptsize $x\ \mathrm{[m]}$}}
\psfrag{ym}{\hspace{0.0mm}\raisebox{0.3mm}{\scriptsize $y\ \mathrm{[m]}$}}
\includegraphics[width=1.2\columnwidth,trim={9mm 25mm 7mm 28mm},clip=true]{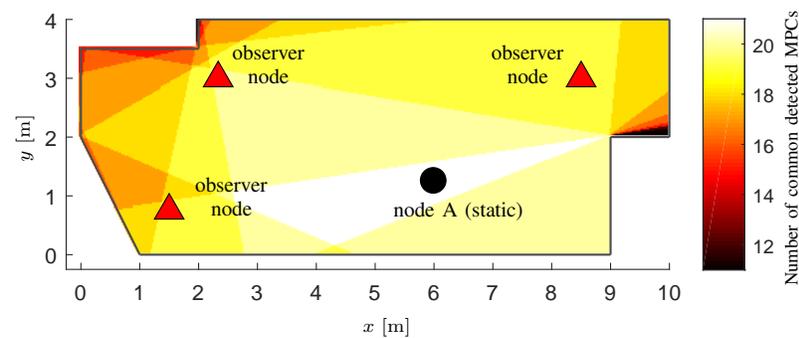}\label{fig:NoMPCMulti}}
\put(-161,47){\footnotesize{node A (static)}}
\put(-242,52){\footnotesize{\begin{tabular}{c}observer\\node\end{tabular}}}
\put(-228,102){\footnotesize{\begin{tabular}{c}observer\\node\end{tabular}}}
\put(-130,102){\footnotesize{\begin{tabular}{c}observer\\node\end{tabular}}}
\put(-13,21){\rotatebox{90}{\scriptsize Number of common detected MPCs}}

\caption{Number of common detected MPCs between the channel(s) from static node A ($\bullet$) to the observer(s) and the channel(s) from mobile node B (anywhere in the room) to the observer(s). The figure is associated with the SINR-based MPC detection criterion described in Sec.~\ref{sec:RealisticSim} with the same floor plan and node setup as in Fig.~\ref{fig:RT}.}
\end{figure*}

\end{document}